\documentclass[conference]{IEEEtran}
\IEEEoverridecommandlockouts

\usepackage[numbers,sort&compress]{natbib} 
\usepackage{amsmath,amssymb,amsfonts}
\usepackage{algorithmic}
\usepackage{graphicx}
\usepackage{textcomp}
\usepackage{xcolor}
\usepackage{comment}
\usepackage{svg}
\usepackage{geometry}
\usepackage[table]{xcolor}
\def\BibTeX{{\rm B\kern-.05em{\sc i\kern-.025em b}\kern-.08em
    T\kern-.1667em\lower.7ex\hbox{E}\kern-.125emX}}

\begin{document}

\title{Can Small GenAI Language Models Rival Large Language Models in Understanding Application Behavior?
}

\author{Mohammad Meymani, Hamed Jelodar, Parisa Hamedi, Roozbeh Razavi-Far, and Ali A. Ghorbani\\
\textit{Canadian Institute for Cybersecurity} \\
\textit{Faculty of Computer Science} \\
\textit{University of New Brunswick} \\
Fredericton, Canada \\
\{mohammad.meymani79, h.jelodar, parisa.hamedi, roozbeh.razavi-far, ghorbani\}@unb.ca}

\maketitle

\begin{abstract}
Generative AI (GenAI) models, particularly large language models (LLMs), have transformed multiple domains, including natural language processing, software analysis, and code understanding. Their ability to analyze and generate code has enabled applications such as source code summarization, behavior analysis, and malware detection. In this study, we systematically evaluate the capabilities of both small and large GenAI language models in understanding application behavior, with a particular focus on malware detection as a representative task. While larger models generally achieve higher overall accuracy, our experiments show that small GenAI models maintain competitive precision and recall, offering substantial advantages in computational efficiency, faster inference, and deployment in resource-constrained environments. We provide a detailed comparison across metrics such as accuracy, precision, recall, and F1-score, highlighting each model’s strengths, limitations, and operational feasibility. Our findings demonstrate that small GenAI models can effectively complement large ones, providing a practical balance between performance and resource efficiency in real-world application behavior analysis.
\end{abstract}


\section{Introduction}
Large language models (LLMs) have noticeably impacted generative artificial intelligence (AI) in recent years, becoming one of the most influential tools in modern machine learning. At their core, LLMs rely on attention mechanisms, first proposed in 2017 \cite{vaswani2017attention,jelodar2025large}, which allow the models to dynamically weigh the importance of different tokens in a sequence. This mechanism enables LLMs to capture long-range dependencies and complex patterns in data, making them highly effective for tasks involving sequential or structured information, such as natural language processing and code understanding. Over time, LLM architectures have evolved in size and complexity. As a result, they can learn from massive datasets and generalize across a wide variety of tasks. Small language models (SLMs), on the other hand, are lightweight, efficient, and fast versions of LLMs, making them suitable for deployment on resource-constrained devices \cite{irugalbandara2024scaling,wang2024comprehensive}. Table \ref{tab:llm-vs-slm} shows the key differences between LLMs and SLMs.

\begin{table}[h]
    \centering
    \caption{Differences between models with respect to different metrics : small language model (SLM), Large Language Model (LLM)}
    \resizebox{\linewidth}{!}{
    \begin{tabular}{l l l}
        \hline
        Metric &  LLM & SLM\\
        \hline
        Inference Speed & Low to moderate & High\\
        Memory Usage & High & Low to moderate\\
        Number of Parameters & 7B-1T & 1B-7B\\
        Training/Tuning Cost & High & Low\\
        Context Length & Long & Short\\
        Maintenance Cost & High & Low\\
        Knowledge Breadth & Extensive & Domain Specific\\
        Energy Consumption & High & Low\\
        Deployment Feasibility & Cloud only & Cloud, Edge/Local machine\\
        \hline
    \end{tabular}}
    \label{tab:llm-vs-slm}
\end{table}

Beyond their architectural innovations, LLMs have demonstrated remarkable capabilities across numerous domains, including mathematics, physics, chemistry, psychology, computer vision (CV), and programming \cite{dong2024survey,jelodar2025large,jelodar2025large2,jelodar2025nld}. Their ability to model complex patterns and generate meaningful outputs has led to breakthroughs in scientific research, automated code generation, educational applications, and AI-assisted problem solving. This wide range of applications has attracted significant research attention, resulting in numerous evaluations, benchmarks, and practical deployments across different fields.

LLMs empower code-related tasks such as code generation, code understanding, malware detection, natural language description (NLD), and so on \cite{jelodar2025large,jelodar2025large2,jelodar2025nld,jelodar2025xgenqexplainabledomainadaptivellm}. In this research, we aim to evaluate LLMs in malware detection, which is formulated as a binary classification problem. The growing threat of malware in software systems has made automated malware detection a critical research area. Traditional signature-based methods struggle to detect novel or obfuscated malware variants, which motivates the use of intelligent models capable of understanding code semantics. LLMs, with their ability to model complex relationships in sequences of tokens, offer a promising solution by learning patterns from large code corpora and effectively identifying malicious behavior. 

Despite their success, the effective use of LLMs in cybersecurity is still challenging because they need a huge amount of computational power. State-of-the-art models usually contain billions of parameters and require strong GPUs, large memory capacity, and notable energy consumption during inference \cite{tu2024overview}. These requirements can limit their deployment in practical security situations, where real-time analysis, scalability, and cost efficiency are essential factors.

Modern malware continuously evolves through obfuscation, polymorphism, and code reuse techniques, making traditional signature-based approaches ineffective. Language models provide a semantic understanding of source code and can identify behavioral patterns that may not be captured through traditional rule-based approaches \cite{canfora2015obfuscation,zhang2023malware}. As a result, evaluating whether SLMs can achieve comparable malware detection performance to larger models is an important research question with direct implications for practical cybersecurity deployment.

In addition to detection performance, understanding the relationship between model's size and operational functionality is important. Security operations centers and edge devices often require rapid inference under constrained computational resources. Therefore, investigating whether smaller models can provide acceptable detection capabilities while reducing latency and resource consumption can strongly impact the future of malware analysis frameworks.

Our contributions are as follows:
\begin{itemize}
    \item We compare different LLMs and SLMs in terms of malware detection, providing a variety of metrics to support the results.
    \item Metrics such as precision, recall, and F1-score are reported for both benign and malware classes, highlighting the ability of the models to distinguish between malware and benign code.
    \item We use 10,000 samples from the SBAN dataset \cite{jelodar2025sban} to make the experiments more reliable.
    \item We compared the results from the classification head of the models as well as direct prompting.
\end{itemize}

In Section \ref{sec:related-works}, we introduce the related works. In Section \ref{sec:methodology}, we explain our methodology. In Section \ref{sec:experiments}, we demonstrate our experimental results. In Section \ref{sec:discussion}, we discuss the limitations and challenges. Finally, in Section \ref{sec:conclusion}, we explore future directions.

\begin{figure*}[h]
    \centering
    \includegraphics[width=\linewidth]{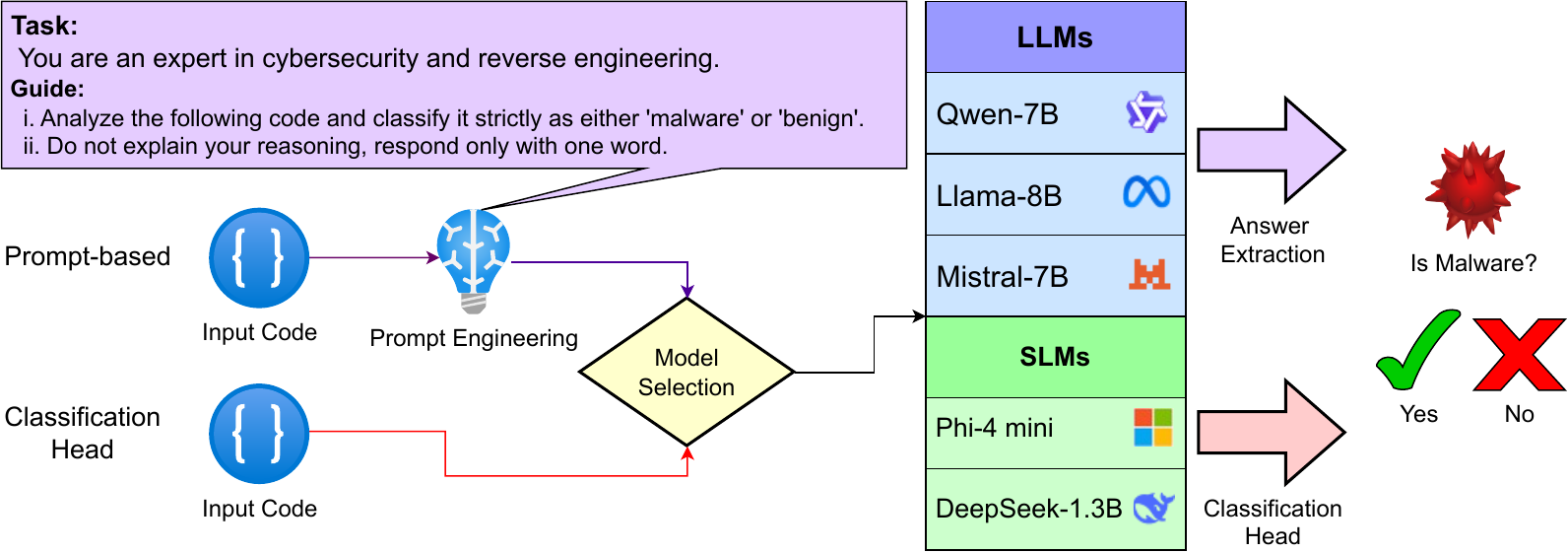}
    \caption{The workflow of our proposed approaches.}
    \label{fig:workflow}
\end{figure*}

\section{Related Works}
\label{sec:related-works}
Recent studies have explored the application of LLMs in malware detection and analysis. For instance, the authors in \cite{fujii2024feasibility} demonstrated the feasibility of using LLMs to support static malware analysis, achieving up to 90.9\% accuracy in explaining malware functionality. In \cite{jelodar2025large}, the authors provided a comprehensive review of LLM applications in software security, covering areas such as code analysis, malware analysis, and reverse engineering. They emphasized the importance of LLMs in understanding code semantics and detecting malicious behaviors. In \cite{marco2025small}, the authors compared SLMs' performance against LLMs and humans in short creative writing and empirically proved that SLMs can rival LLMs in specific contexts.

Although LLMs have achieved remarkable performance, their deployment often requires substantial computational resources. Consequently, SLMs have emerged as a promising alternative for practical cybersecurity applications. Recent studies suggest that carefully optimized SLMs can achieve competitive performance on domain-specific tasks while reducing inference latency, memory usage, and energy consumption \cite{taylar2026strategies}. This makes them particularly attractive for edge computing environments, embedded security systems, and large-scale malware scanning infrastructures where computational efficiency is a critical requirement.

To the best of our knowledge, this study is the first to compare the performance of various LLMs (SLM vs LLM) in malware code diagnosis, utilizing both their classification heads and direct prompting, with the goal of understanding application behavior through language models.

\section{Methodology}
\label{sec:methodology}
In this Section, we first explain the procedure of data collection and LLM selection. Then, we describe the classification methodology. Finally, we introduce evaluation metrics. Figure \ref{fig:workflow} shows a simple workflow of both approaches.

\subsection{Data Collection}
We use SBAN \cite{jelodar2025sban} as a benchmark, which is a multi-dimensional dataset including both benign and malware source code from both publicly available sources and AI-generated code. SBAN contains more than 3 million samples, including both benign and malware code.

The SBAN dataset was selected because it provides a diverse collection of samples covering both benign and malicious source code. The dataset includes code originating from open-source repositories, security benchmarks, and AI-generated codes. Such diversity reduces dataset bias and allows a more realistic evaluation of language models in practical software security scenarios.

To ensure balanced evaluation, we randomly selected 5,000 benign and 5,000 malware samples. A balanced dataset prevents accuracy inflation caused by class imbalance and allows precision, recall, and F1-score to better reflect actual model behavior. Since malware detection often suffers from skewed distributions in real-world datasets, balanced evaluation provides a controlled environment for comparing model capabilities.
\subsection{Language Model Selection}
To test the models, we use DeepSeek, Phi, Llama, Qwen, and Mistral. For implementation, we use \texttt{HuggingFace} and  \texttt{Transformers} libraries from Python to load the models by their ids. These ids and their corresponding abbreviations are shown at Table \ref{tab:abbre}.

The selected models represent both SLM and LLM categories, and are chosen based on their popularity, availability, and demonstrated performance on code-related tasks. DeepSeek-Coder and Phi-4-mini represent lightweight models with relatively low computational requirements, while Llama-3.1-8B, Qwen-2.5-Coder-7B, and Mistral-7B represent larger instruction-tuned models capable of advanced reasoning and code understanding. This diversity allows us to evaluate how model's size influences malware detection performance.
\begin{table*}[h]
    \centering
    \caption{Models' id and abbreviation.}
    \resizebox{0.75\linewidth}{!}{
    \begin{tabular}{l l l}
        \hline
        id & abbreviation & Type\\
        \hline
        \texttt{deepseek-ai/deepseek-coder-1.3b-instruct} & DeepSeek-1.3B&Small\\
        \texttt{microsoft/Phi-4-mini-instruct} & Phi-4-mini&Small\\
        \texttt{meta-llama/Llama-3.1-8B-Instruct}&Llama-3.1-8B&Large\\
        \texttt{Qwen/Qwen-2.5-Coder-7B-Instruct}&Qwen-2.5-7B&Large\\
        \texttt{mistralai/Mistral-7B-Instruct-v0.3}&Mistral-7B&Large\\
        \hline
    \end{tabular}}    
    \label{tab:abbre}
\end{table*}
\subsection{Methodology}
In this section, we explain the classification methodology for both classification head and prompt-based strategies.
\subsubsection{Classification Head}
Transformer-based models produce embeddings based on the context, which can be used for different classification tasks by adding lightweight classification heads. The probability of each class is obtained via the sigmoid function:
\begin{equation}
p(y=1 \mid x) = \sigma(z) = \frac{1}{1 + e^{-z}}    
\end{equation}
where $x$ is the input sequence, $y \in \{0,1\}$ is the label, and $z$ denotes the logits produced by the model. The default threshold is 0.5, where the values below that amount are classified as zero, otherwise as one.
\subsubsection{Prompt-based}
The weights associated with the classification head are not well optimized for malware classification tasks. Therefore, we use zero-shot prompting by directly asking the model to determine whether the code is malware. Figure \ref{fig:workflow} shows the prompt we used during answer generation. After receiving the result, we need to process the generated answer to extract the final result from it. We did this by simply trimming the final result and searching for the keywords.
\begin{figure*}[h]
    \centering
    \includegraphics[width = \linewidth]{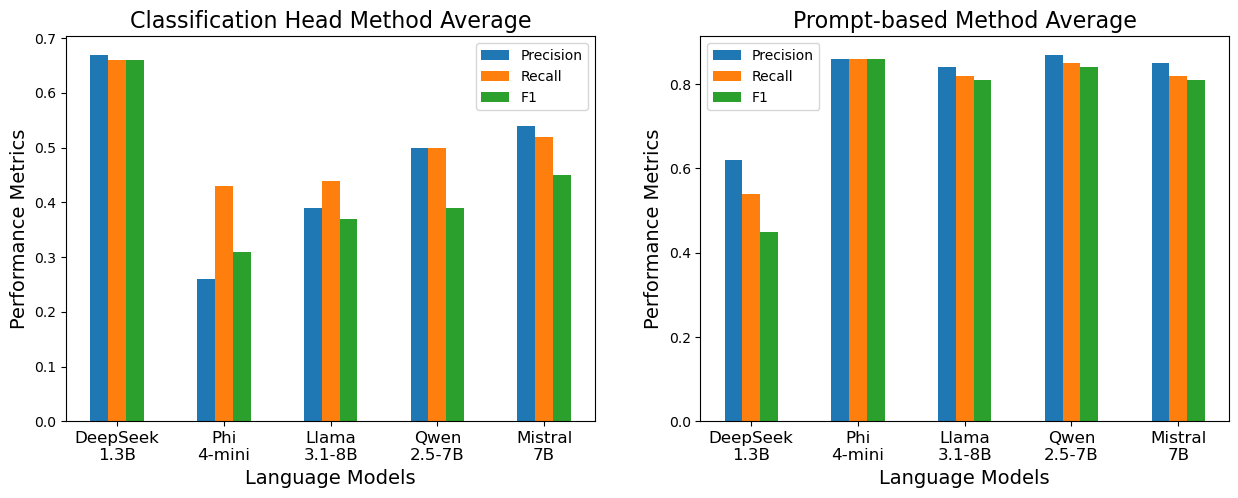}
    \caption{Grouped bar chart comparing weighted and macro averages of precision, recall, and F1-score across both classes for each model. Results are obtained using a subset of 10,000 samples from the SBAN dataset (5,000 benign and 5,000 malware). Models are evaluated under both classification head and prompt-based strategies, with default classification threshold set to 0.5 and prompt temperature set to 0.}
    \label{fig:bar-comparison}
\end{figure*}

\subsection{Evaluation and Validation}
As previously mentioned, we label 0 as benign and 1 as malware. To evaluate the performance of the trained models, we use metrics such as accuracy, precision, recall, F1-score, and support. Since we assess the LLMs on a binary classification problem, these metrics are provided for both classes.

Here, TP denotes true positives, FN denotes false negatives, FP denotes false positives, and TN denotes true negatives. Then, the formulas for accuracy, precision, and recall are as follows:
\begin{align}
&\text{Accuracy}  = \frac{TP + TN}{Total} \\[1em]
&\text{Precision} = \frac{TP}{TP + FP} \\[1em]
&\text{Recall}    = \frac{TP}{TP + FN} \\[1em]
&\text{F1}  = \frac{2 \times \text{Precision} \times \text{Recall}}{\text{Precision}+\text{Recall}}
\end{align}

Accuracy provides an overall measure of classification correctness but cannot fully capture the model's effectiveness in security applications. In malware detection, false negatives can have severe consequences because malicious samples may evade detection. Therefore, precision and recall become particularly important metrics. High precision indicates that detected malware samples are likely to be truly malicious, while high recall demonstrates the model's ability to identify the majority of malware instances. The F1-score combines both metrics and provides a balanced assessment of detection capability.

\section{Experiments}
\label{sec:experiments}
In this section, we present the experimental setup, training configurations, hardware environment, and detailed evaluation of all models tested on the malware detection task. We conducted extensive experiments to compare the performance of multiple LLMs and highlight the trade-offs between accuracy, speed, and computational cost.

\subsection{Training Setting and Hyperparameters}
In this section, we present the hyperparameters used during evaluation.
\subsubsection{Classification Head}
All models were evaluated using instruction-tuned variants designed for code understanding and classification tasks. For each model, we followed the recommended hyper parameters provided in their respective documentation or papers. Specifically, a batch size of 8 sequences per GPU was used, with a learning rate of $1e^{-5}$ and a linear warmup schedule over the first 100 steps. Each model was evaluated over a single inference pass, as pre-trained weights were employed for classification. The maximum sequence length was set to 1024 tokens, and optimization was performed using the Adam optimizer with a weight decay of 0.01.
\subsubsection{Prompt-based}
In the prompt-based strategy, we used the same prompt for all the models, assigning the temperature to zero. These settings ensured a fair comparison across all models, balancing computational feasibility with sufficient context for malware detection.
\begin{table*}[h]
    \centering
    \caption{Performance metrics for each LLM on malware detection. Class 0 denotes benign samples, Class 1 denotes malware samples.}
    \begin{tabular}{l | c | c c | c c | c c}
         \hline
         Model ID & \cellcolor{green!60}Accuracy &\multicolumn{2}{|c|}{\cellcolor{green!50}Precision}& \multicolumn{2}{c}{\cellcolor{green!40}Recall}& \multicolumn{2}{|c}{\cellcolor{green!30}F1-score}\\
         \hline
         \cellcolor{gray!40}Classification Head Method&&0&1&0&1&0&1\\
         \hline
         DeepSeek-1.3B&66\%&72\%&63\%&53\%&79\%&61\%&70\%\\
         Phi-4-mini&43\%&6\%&46\%&1\%&84\%&2\%&60\%\\
         Llama-3.1-8B&44\%&32\%&47\%&10\%&79\%&15\%&59\%\\
         Qwen-2.5-7B&50\%&50\%&50\%&93\%&7\%&65\%&12\%\\
         Mistral-7B&52\%&51\%&56\%&88\%&16\%&64\%&25\%\\
         \hline
         \cellcolor{gray!40}Prompt-based Method&&0&1&0&1&0&1\\
         \hline
         DeepSeek-1.3B&54\%&72\%&52\%&14\%&94\%&23\%&67\%\\
         Phi-4-mini&86\%&90\%&83\%&81\%&91\%&85\%&87\%\\
         Llama-3.1-8B&82\%&93\%&75\%&69\%&95\%&79\%&84\%\\
         Qwen-2.5-7B&85\%&78\%&96\%&97\%&73\%&86\%&83\%\\
         Mistral-7B&82\%&74\%&97\%&98\%&66\%&84\%&79\%\\
         \hline
    \end{tabular}
    \label{tab:results}
\end{table*}
\subsection{Hardware and GPU Characteristics}
All experiments were conducted on a workstation equipped with a single NVIDIA H100 GPU, 80 GB of VRAM, and 512 GB RAM. This hardware provided sufficient computational power to load and run large LLMs such as Qwen-7B and Mistral-7B, efficiently. Models were tested sequentially to avoid GPU memory bottlenecks, and inference times were recorded for each model.
\begin{figure*}[h]
    \centering
    \includegraphics[width = \linewidth]{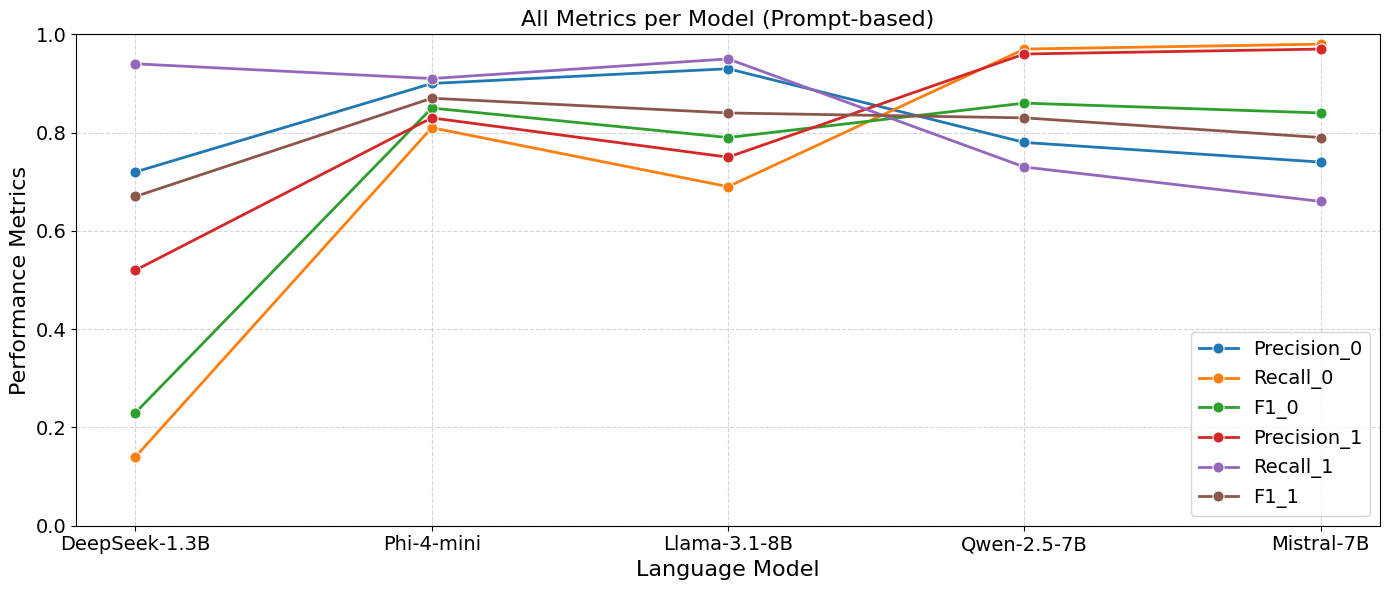}
    \caption{Line charts showing the variation of performance metrics. Each chart focuses on a single metric for clearer model performance comparison.}
    \label{fig:metric-charts}
\end{figure*}

\begin{figure}[h]
    \centering
    \includegraphics[width = \linewidth]{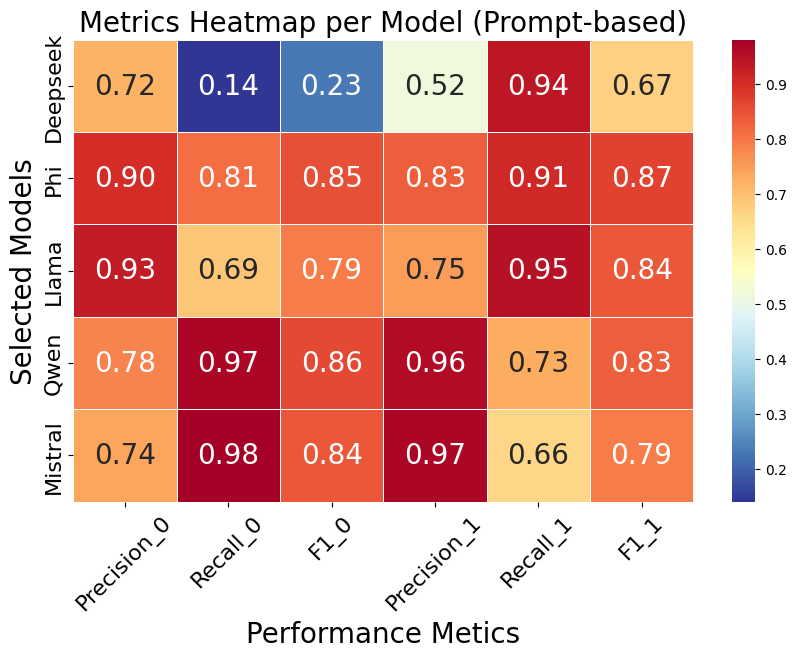}
    \caption{Heatmap visualization of precision, recall, and F1-score for both classes across all evaluated models using the prompt-based method. Results are based on 10,000 samples from the SBAN dataset with equal class distribution. Warmer colors indicate higher performance. Evaluation is conducted under consistent settings, including zero-shot prompting and temperature set to 0.}
    \label{fig:heatmap}
\end{figure}

\subsection{Results and Analysis}
We denote\textit{ Class 0} as benign and \textit{Class 1} as malware samples. Metrics include Accuracy, Precision, Recall, F1-score, and inference time. 

To ensure reproducibility and clarity in performance comparisons, all visualizations (Figures \ref{fig:metric-charts}, \ref{fig:bar-comparison}, and \ref{fig:heatmap}) are generated using the same experimental setup. Specifically, we evaluate all models on a balanced subset of 10,000 samples from the SBAN dataset, consisting of 5,000 benign and 5,000 malware instances. For classification head experiments, a threshold of 0.5 is used to determine class labels. In the prompt-based setting, zero-shot prompting is applied with temperature set to 0 to ensure deterministic outputs. All metrics are computed separately for each class.

Figure \ref{fig:bar-comparison} presents the weighted average of the metrics for both classes across all models. Figure \ref{fig:metric-charts} illustrates line plots for precision, recall, and F1-score. These plots highlight detailed comparisons between models. Figure \ref{fig:heatmap} provides alternative visualizations of precision, recall, and F1-score for more granular insights.

An interesting observation is the significant performance gap between classification-head and prompt-based approaches. Across nearly all evaluated models, prompt-based inference consistently outperformed the classification-head strategy. This suggests that the instruction-tuning objectives used during model training enable stronger reasoning capabilities than the default classification layers.

Another notable finding is the performance of Phi-4-mini. Despite being considerably smaller than several competing models, it achieved one of the highest F1-scores and maintained balanced precision and recall values. This result supports the hypothesis that model size alone does not determine malware detection effectiveness. Instead, training quality, instruction tuning, and domain knowledge may play equally important roles.

Furthermore, larger models such as Qwen-2.5-Coder and Mistral-7B demonstrated strong malware detection capabilities but required substantially greater inference time. This trade-off highlights an important practical consideration: organizations must balance detection performance against computational cost and deployment constraints. However, the computational cost and inference time must be considered when selecting a model for deployment in real-world security applications. The visualization of metrics allows for more informed model selection based on specific operational needs, such as prioritizing recall over speed for malware prevention systems.

From a practical deployment perspective, the results indicate that SLMs may serve as an effective first-stage filtering mechanism in malware analysis pipelines. Suspicious samples identified by SLMs could subsequently be forwarded to larger models for deeper analysis and explanation. Such hierarchical architectures could reduce operational costs while preserving high detection accuracy.

Additionally, the strong performance achieved by prompt-based methods suggests that extensive task-specific retraining may not always be necessary. Organizations can use pre-trained language models with carefully designed prompts to perform malware detection tasks with minimal additional training cost.

\section{Discussion and Limitations}
\label{sec:discussion}

This study highlights the potential of SLMs for malware detection, emphasizing operational efficiency and feasibility in resource-constrained environments. While very large models generally achieve higher overall accuracy, SLMs such as DeepSeek-Coder and Phi-4-mini offer competitive performance with significantly lower computational requirements and faster inference times. This makes them particularly suitable for real-world deployment, where processing large codebases in an efficient manner is critical.

Our experiments show that SLMs perform reliably on malware samples, maintaining high precision, but exhibit lower recall for benign. Despite this limitation, the results demonstrate that carefully designed SLMs can provide meaningful detection capabilities, highlighting their practical value for automated malware analysis. Visualizations such as heatmaps and grouped bar charts effectively illustrate model strengths and weaknesses across metrics, providing clearer insights into model behavior and potential areas for improvement.

Several factors may influence the generalizability of the reported results. First, the experiments were conducted on a subset of the SBAN dataset rather than the complete corpus. Although the selected samples were balanced across malware and benign classes, performance may vary when evaluated on larger or more diverse datasets. Second, only five language models were evaluated, and additional models may show different behaviors. Third, the prompt-based experiments relied on a single prompt template. Alternative prompt engineering strategies may further improve performance. Finally, the evaluation focused exclusively on static source code analysis and did not consider dynamic behavioral information, which could provide additional context for malware detection.

Future work will focus on enhancing the performance of SLMs through strategies such as fine-tuning on diverse malware datasets and hybrid static-dynamic analysis. Additionally, model optimization methods like pruning, distillation, and compression will be explored to improve inference speed, while preserving detection capability \cite{liu2022fewfine, ajayi2024exploring, feng2025llm, kumar2024prompt}. These refinements aim to make SLMs robust and practical for real-time malware detection, demonstrating that smaller models can be a viable alternative to very large models in operational contexts.

\section{Conclusion}
\label{sec:conclusion}

In this study, we systematically evaluated both small and large language models for understanding application behavior and detecting malware code, considering metrics such as accuracy, precision, recall, and F1-score. While larger models generally achieved higher overall accuracy, smaller models like Phi-4-mini demonstrated competitive performance and strong reliability across both benign and malware samples.

These results highlight that small language models can offer practical, resource-efficient solutions, particularly in environments with limited computational capacity. Our findings provide guidance for selecting and optimizing models for application behavior analysis and point to future directions such as fine-tuning, hybrid analysis approaches, and model compression to further improve performance.

\bibliographystyle{ieeetr}
\bibliography{main}

\end{document}